\documentstyle[editedvolume]{crckapb} 

\def\mM{\ifmmode(m{-}M)\else$(m{-}M)$\fi}
\def\vi{\ifmmode(V{-}I)\else$(V{-}I)$\fi}
\def\viz{\ifmmode(V{-}I)_0\else$(V{-}I)_0$\fi}

\def\mbar{\ifmmode\overline m\else$\overline m$\fi}
\def\mibar{\ifmmode\overline m_I\else$\overline m_I$\fi}
\def\Mbar{\ifmmode\overline M\else$\overline M$\fi}
\def\MIbar{\ifmmode\overline M_I\else$\overline M_I$\fi}
\def\MVbar{\ifmmode\overline M_V\else$\overline M_V$\fi}
\def\MRbar{\ifmmode\overline M_R\else$\overline M_R$\fi}
\def\MKbar{\ifmmode\overline M_K\else$\overline M_K$\fi}
\def\MHbar{\ifmmode\overline M_H\else$\overline M_H$\fi}
\def\mrbar{\ifmmode\overline m_R\else$\overline m_R$\fi}
\def\mkbar{\ifmmode\overline m_K\else$\overline m_K$\fi}
\def\mvbar{\ifmmode{\overline m_V}\else{$\overline m_V$}\fi}
\def\mfw{\ifmmode{\overline m_{814W}}\else{$\overline m_{814W}$}\fi}
\def\MFW{\ifmmode{\overline M_{814W}}\else{$\overline M_{814W}$}\fi}

\def\etal{{\it et al.}}
\def\hst{{\it HST}}
\def\avemi{\ifmmode\langle\overline{m}_I^0\rangle\else$\langle\overline{m}_I^0\rangle$\fi}

\def\farcs{\hbox{$.\!\!^{\prime\prime}$}}
\def\aj{{\it Astron.~J.}}
\def\apj{{\it Ap.~J.}}
\def\apjl{{\it Ap.~J.\ Lett.}}
\def\apjs{{\it Ap.~J.\ Suppl.}}
\def\mnras{{\it MNRAS}}

\def\pasp{{\it Publ. Astron. Soc. Pac.}}
\def\aap{{\it Astron. Astrophys.}}
\def\aapl{{\it Astron. Astrophys. Lett.}}

\def\lta{\mathrel{\rlap{\lower 3pt\hbox{$\mathchar"218$}}
     \raise 2.0pt\hbox{$\mathchar"13C$}}}
\def\gta{\mathrel{\rlap{\lower 3pt\hbox{$\mathchar"218$}}
     \raise 2.0pt\hbox{$\mathchar"13E$}}}

\begin{opening}
\title{Distances from Surface Brightness Fluctuations}

\author{John P. Blakeslee}
\institute{California Institute of Technology,
	M.S.~105-24, \\ Pasadena, CA 91125;~ 
	jpb@astro.caltech.edu}

\author{Edward A. Ajhar}
\institute{National Optical Astronomy Observatories,
	P.O. Box 26732, \\
	Tucson, AZ 85726;~ 
	ajhar@noao.edu}

\author{John L. Tonry}
\institute{University of Hawaii, 
	2680 Woodlawn Drive, \\
	Honolulu, HI 96822;~ 
	jt@avidya.ifa.hawaii.edu}

\end{opening}
\runningtitle{Surface Brightness Fluctuations}

\begin{document}

\section{Introduction}

The practice of determining galaxy distances from the amplitude
of their spatial fluctuations in surface brightness 
began  with the work of Tonry \& Schneider (1988),
who developed the method in detail and made its
initial application to estimating elliptical galaxy distances.
While several articles over the past decade have included 
some review material
(e.g., Jacoby \etal\ 1992; Tonry 1996; Tonry \etal\ 1997, hereafter SBF-I),
this is the first intended as a comprehensive review
of the surface brightness fluctuation (SBF) method.

The SBF method is conceptually quite simple, the basic idea being
that nearby (but unresolved) star clusters and galaxies appear 
``bumpy,'' while more distant ones appear smooth.  
This is quantified via a measurement of the
Poisson fluctuations in the number of unresolved stars (in an image
of an elliptical galaxy, for instance) encompassed by a CCD pixel.
If $N$ is the mean number of stars per pixel and $\overline f$ is the
mean flux per star, then the mean pixel intensity is $N\overline f$
and the variance is ${N}\overline f^2$. 
Dividing the observed variance by the observed mean yields $\overline f$,
which decreases inversely with the square of the distance ($d^{-2}$).
If the corresponding luminosity $\overline L$ happens to be a standard
candle, the distance follows directly: 
$\,d^2 = \overline L/4\pi\overline f$.

In practice, the measurement is complicated by the fact that adjacent
pixels in real CCD images are correlated through convolution with the
point spread function (psf) due to the atmosphere and telescope optics.
One must therefore determine the variance from the amplitude of the
image power spectrum on the scale of the psf.  As only celestial
sources suffer psf-blurring, this is a mixed blessing, making it
possible to remove all noise which does not favor this scale.
Examples include noise due to photon counting statistics and 
CCD read noise.

A further complication is that $\overline L$ is not by itself 
a standard candle in optical bandpasses.  However, both theory and
observation indicate that it varies in a simple, predictable manner
among old stellar populations.  The most widely used version of
the SBF method employs a standard candle constructed from a linear combination
of \MIbar\ and broad-band \vi\ color (e.g., Tonry 1991; SBF-I;
Ajhar \etal\ 1997), where \MIbar\ is the absolute magnitude of
$\overline L$ in the Kron-Cousins $I$-band.

Being the ratio of the second and first moments of the stellar luminosity
function, $\overline L$ is the luminosity-weighted average
stellar luminosity \cite{TS88}.  It is therefore weighted towards the
brightest stars in a population; for evolved populations, these
are red giant branch (RGB) stars.  Since RGB stars are red,
SBF magnitudes are red; $\mvbar{-}\mibar\approx2.4$ 
is a typical SBF color for an elliptical galaxy 
(Tonry \etal\ 1990, hereafter TAL90).
It was for this reason, as well as its relative insensitivity
to stellar population differences, that \mibar\ was 
the magnitude of choice for
the ``SBF~Survey of Galaxy Distances'' \cite{sbfI}.

The following section describes the technical details and
difficulties involved in the SBF method, primarily
as applied in the $I$-band SBF Survey but including
approaches used by other authors.  Section~3 then discusses the
theoretical calibration of the method and various stellar population
effects.  Details on the data sample, empirical calibration,
and results of the SBF Survey are given in Section~4, including 
comparisons with other methods and constraints on the Hubble constant.
Sections 5--6 describe other optical and near-infrared
SBF distances measured from the ground, and
Section~7 discusses the recent and exciting results
with the {\it Hubble Space Telescope} (\hst).
Section~8 evaluates the strengths, shortcomings, universality,
and applicability of the SBF method, and the final
section forecasts what the future may hold for it.

\section{Measuring the Surface Brightness Fluctuations}

Ajhar \etal\ (1998) give a complete presentation of SBF theory and
the analysis techniques used in the ground-based $I$-band survey.
Here, we summarize the method and refer the reader to
that work for further details.

The ultimate goal for a galaxy image suitable for SBF measurement is
that the pixel-to-pixel fluctuations be dominated by the galaxy's
stellar Poisson statistics rather than by flattening errors, CCD
artifacts, photon statistics, {\it etc}.  Consequently,
successful SBF observations require
extraordinary care and planning.  Typically, sufficient CCD
calibration requires bias and dark frames, high signal-to-noise 
(S/N) flat fields, and, for thinned CCDs, high signal-to-noise 
fringe frames.  Even ``small'' amplitude fringing of $\lta\,$1\% must
be removed for accurate fluctuation measurements because
it can have significant power on the scale of the psf.
We have found fringe patterns to be remarkably stable and removable
from galaxy images by subtracting the fringe frame,
appropriately scaled to the observation.

Under nominal observing conditions, the exposure time on a galaxy is
dictated by (a) the SBF amplitude \mbar\ (based on its expected
distance) and (b) the sensitivity of the detector.  Fundamentally, the
exposure time must be long enough so that the photon shot noise per
pixel is less than the stellar surface brightness fluctuations.  The
approximate break-even point occurs when one photon is collected per
giant star of brightness \mbar.  In this way, the exposure time is
given by
\begin{equation}
t \,=\, {S \over N} 10^{0.4(\mbar - m_1)},
\end{equation}
where $m_1$ is the magnitude yielding 1~detected photo\-electron 
per second,  
and $S/N$ is the desired signal-to-noise ratio.  One
normally strives to collect $\sim 5$--10~$e^-$ per \mbar\ star (or
$S/N \sim 5$--10).  The general observation procedure is to take 3--10
exposures of 300--600$\,$s each, dithering each exposure by
$\sim\,$5$^{\prime\prime}$
perpendicular to the parallel clock direction of the CCD
to improve flattening and the removal of fringe patterns and CCD defects.

Because the $I$-band SBF calibration is sensitive to the \vi\ color of
the galaxy, precise color measurements are required for a reliable
distance measurement.  As a result, sufficient time must be spent
observing standard stars to ensure a precise photometric calibration.

Once the data are collected and the initial CCD calibration is
complete (including fringe removal, if necessary), we register and
stack the series of $I$-band galaxy observations and remove cosmic ray
events.  The final $I$-band image is used with a $V$-band image,
usually observed at the same time, to determine the \vi\ color of the
galaxy.  We first mask obvious point sources, background galaxies, and
any dusty regions in the galaxy. 
We estimate the sky levels by fitting the outer regions of the galaxy
to an $r^{1/4}$ law profile plus a constant sky offset.  Based on
these sky estimates, we compute \vi\ colors for the galaxy as a
function of radius, corresponding to the \mbar\ measurements to be made.

The next step is to fit and remove the galaxy.  
First, we mask all the obvious objects and any dust.
Next, we fit elliptical isophotes to the sky-subtracted galaxy image
and build a model of the galaxy pixel-by-pixel by
interpolating between fitted ellipses.  Finally, we subtract the model
from the data image.
The residual large-scale deviations from a flat background 
are then fitted and subtracted from the image.  We note
that this technique does corrupt the lowest wavenumbers of the
image power spectrum.  In addition, imperfect flattening and 
fringe removal potentially corrupt the low wavenumbers
as low spatial frequency power
may be introduced into the data at that stage of the reduction.
However, this is irrelevant in the end because we omit low
wavenumbers from the determination of \mbar.

While some authors \cite{PM94} have used a similar method to remove
the mean galaxy light, others have employed some kind of adaptive
filter.  Lorenz \etal\ (1993) used a Laplace filter to subtract galaxy
light from the image.  Neilsen \etal\ (1997) employed a Butterworth
filter in Fourier space to remove the low-wavenumber galaxy component
from the power spectrum.  These methods circumvent the need for a
galaxy model and may have advantages in measuring \mbar\ for galaxies
whose structure makes them difficult to fit.  We note that the overall
approaches of Lorenz \etal\ (1993), Neilsen \etal\ (1997), and
Sodemann \& Thomsen (1995) in determining \mbar\ is slightly different
from the presentation here.

After the mean galaxy profile is removed, the next step is to identify
foreground stars, background galaxies, and globular clusters (GCs) in the
image.  We use a modified version of {\sc DoPhot} \cite{SMS93} to
catalogue all objects in the image.  Next, we characterize
the distribution of these objects in magnitude and radius.
This step is very important because the fluctuation
amplitude that we measure $P_0$ includes a {\em residual\/}
fluctuation signal $P_r$ from {\em undetected\/} faint
GCs and background galaxies.  Ultimately, we want to
measure the net fluctuation signal
\begin{equation}
P_{\rm fluc} \,=\; P_0 - P_r\,.  \label{eq:pfluc}
\end{equation}
For galaxies at large distances where only the brightest GCs are
detectable, $P_r$ swamps $P_{\rm fluc}$.  The SBF
technique then becomes a powerful tool for studying the GCs of distant
galaxies (Blakeslee \& Tonry 1995; Blakeslee \etal\ 1997).  
To estimate $P_r$, we assume that the globular cluster luminosity
function is Gaussian and that the 
background galaxy luminosity function 
is a power law.  Fortunately, except when the
data are marginal, even a generous
error allowance in this step typically contributes only a small amount
to the final uncertainty in \mbar.  As a result, any given
measurement of \mbar\ is relatively insensitive to the details of 
these~assumptions.~ 

Next, we determine the total fluctuation amplitude $P_0$.  After
choosing a suitable psf from the galaxy image, we build an expectation
power spectrum $E(\vec k)$ from the psf, the smooth galaxy model, and
a mask.  (The mask selects the region of the galaxy to be
analyzed and excludes GCs, galaxies, and dust.)  The
power spectrum of the masked data $P(\vec k)$ is computed and fitted
to the expectation power spectrum such that
\begin{equation}
P(\vec k) \;=\; P_0\,E(\vec k) \,+\, P_1\,,  \label{eq:pk}
\end{equation}
where $P_1$ is a flat ``white noise'' component.
The ratio $\,(P_0{-}P_r)/P1 = P_{\rm fluc}/P1$ is then another
good indicator of the signal-to-noise level.
Finally, we compute $\,\mbar = -2.5 \log(P_{\rm fluc}/t) + m_1$.

\section{Theoretical Calibrations of the SBF Method}

\subsection{Initial Efforts}

TAL90 made a pioneering attempt
at theoretically calibrating SBF magnitudes.  
They used the Revised Yale Isochrones
(RYI, Green \etal\ 1987) to~calculate \MVbar, \MRbar, and \MIbar\ for
model stellar populations covering the conceivable range in metallicity,
age, initial mass function, and helium abundance.
They also allowed for red or blue horizontal branches
and a possible asymptotic giant branch.
These RYI-based models indicated that \MVbar\ and \MRbar\
were highly sensitive to population age and metallicity, becoming
sharply fainter as the population became redder.

However, \MIbar\ for these models showed very little variation.
At the metallicities of elliptical galaxies, it actually
became a bit brighter in redder populations.  The purely
theoretical RYI calibration, quoted here solely for historical
purposes, corresponded to:
$\; \MIbar \,=\, -1.93 \,-\, [\viz{-}1.15].$ 
This was a bad calibration;
blind acceptance of it would haved yielded a Virgo cluster
distance of 21~Mpc.
However, TAL90 realized that at least the zero~point 
was wrong and so made it fainter by 0.4~mag to be in accord with an
assumed distance of 0.7~Mpc for M32.  What they were unable to
determine from the available data was that the slope of the relation
was also in serious error.  By purely empirical means, Tonry (1991)
found that the correct slope was several times larger and had the
opposite sign, with \MIbar\ fainter in redder populations.

A new edition of the Yale Isochrones, employing much improved
model atmosphere flux curves is now available (Demarque \etal\ 1996).
These appear to predict fluctuation magnitudes consistent with
observations and with the models discussed in the following section
(S.~Yi 1996, priv.\ comm.).  However, to date no theoretical
SBF magnitudes based on these new iso\-chrones have been published.

\subsection{The Current Theoretical Calibration}

Worthey (1993a, 1994) has produced a realistic set of 
stellar population models employing some of the best available
stellar atmospheric and evolutionary models.  He has calculated
SBF magnitudes for all the standard optical and near-infrared
bandpasses.  These calculations (as well as those of
Buzzoni 1993) confirm that \MIbar\ is the most favorable of 
the optical SBF magnitudes for measuring extragalactic distances.

A fit to the Worthey models over the color range appropriate
to elliptical galaxies, $1.05\lta \vi \lta 1.35$, yields
the following theoretical calibration: 
 \begin{equation}
   \MIbar \;=\; -1.83 \,+\, 4.3\,[\viz - 1.15] \;,  \label{eq:worcal}
 \end{equation}
with an rms scatter of 0.10~mag.  The models represent homogeneous
single-burst stellar populations, but age and metallicity
are completely degenerate in their effect here (and changes in
the IMF have very little effect).  Thus, composite stellar
populations will follow this same relation.

Adopting Eq.~(\ref{eq:worcal}) as a theoretical calibration 
makes SBF a ``secondary'' distance indicator, similar to Cepheids.
The predicted intrinsic scatter is 0.10~mag
or less, depending upon how much variation there is among the
stellar populations of elliptical galaxies.  We will see in the
following section that Eq.~(\ref{eq:worcal}) agrees with the latest
empirical calibration to better than 0.1~mag in zero~point.

Buzzoni (1993) has also calculated SBF magnitudes in the
optical and near-infrared, comparing them with observations
from TAL90 and Tonry (1991). 
Since his models were on the Johnson system, he transformed
the observations using fairly large and uncertain corrections
($\sim\,$0.6~mag for \mibar).  Nevertheless, the trend he found
between \MIbar\ and \viz\ was consistent with the
empirical one and the one predicted by the Worthey models.  He
derived a distance to M31/M32 consistent with the Cepheid distance.

\subsection{Stellar Populations Issues \label{ssec:pops}}

A complete discussion of the ramifications of 
SBF magnitudes and colors for the study of stellar populations
is worth a review in its own right.  As this article is dedicated
to the use of SBF as a distance indicator, we here provide only
a cursory treatment of stellar population issues.

TAL90 were the first to consider the effects of age and metallicity
on SBF magnitudes, but they were hampered by inadequate models.
Worthey (1993a) and Buzzoni (1993) both discussed the effects
of stellar population differences on \Mbar\ in various bands,
arriving at somewhat discordant conclusions.  Buzzoni concluded
that SBF colors ``give so far the more direct and confident evidence
of a metal-poor stellar component in [otherwise metal-rich] 
elliptical galaxies,''  while Worthey found ``no evidence for composite
populations in elliptical galaxies on the basis of fluctuation colors.''

In a separate work, Worthey (1993b) showed how SBF 
can be a powerful tool in helping to solve the problem
of the ultraviolet excess in elliptical galaxies.  He 
concluded that ultraviolet SBF magnitudes can easily distinguish
between the presence of young stars and a hot horizontal branch 
in a stellar population.  His proposal has yet to be applied
in practice, however.~~ 

Ajhar \& Tonry (1994) made the first major observational 
study of SBF magnitudes as stellar population gauges.
They measured $(\mvbar{-}\mibar)$ for 19 Galactic globular clusters,
clearly showing a correlation of this SBF color with metallicity,
due primarily at low metallicity to variation in \MVbar.  At the
high metallicity end, these authors surmised, \MVbar\ reaches 
a minimum brightness near the point where the Mg$_2$ spectroscopic
index saturates (Mg$_2{\,\sim\,}0.32\;$mag) in giant ellipticals. 
\MIbar\ continues to grow fainter, however, causing a large
spread of $(\mvbar{-}\mibar)$ at a fixed value of Mg$_2$.

SBF magnitudes also track stellar population gradients within
galaxies.  Tonry (1991) showed that the strong radial gradient
in \mibar\ observed in the dwarf galaxy NGC~205 followed the
\vi\ color gradient.
However, due to the relatively 
recent star formation in this galaxy, the \mibar\
gradient is actually shallower than would be predicted based on
the color gradient \cite{sbfI}.  Sodemann and Thomsen (1995, 1996) have
found SBF magnitude gradients in accord with theoretical expectations
from color gradients in the elliptical galaxies M32 and NGC~3379.

Very little has been published on observations of SBF
blueward of the $V$~band.  Shopbell \etal\ (1993) reported 
$(\overline m_B{-}\overline m_R) = 2.5{\,\pm\,}0.8$ for NGC 5128.
Sodemann \& Thomsen (1996) have measured $\overline m_B$ for M32;
their result combined with $\overline m_R$ from TAL90 yields
$(\overline m_B{-}\overline m_R) = 2.42{\,\pm\,}0.12$. 
The Worthey models predict
$(\overline m_B{-}\overline m_R)$ values in the range of 2.40--2.95 mag
for [Fe/H]${\,\geq\,}{-}1.5$ and $t{\,\geq\,}3\;$Gyr.
We have further, unpublished measurements of this SBF color
for M31 and its dE companions and find fair consistency
with the models, although the data tend toward the blue side.

Jensen \etal\ (1998a) compare the observed plots of
\MKbar\ vs.\ integrated \viz\ to the predictions from
the Worthey models.  The models indicate that more luminous,
redder elliptical galaxies have both higher metallicities and 
greater ages.  The bluer, more compact ellipticals appear to
have both lower mean metallicities and smaller ages, 
reducing the scatter in \MKbar.

\begin{figure}
\vspace{9.5cm}  
\includegraphics{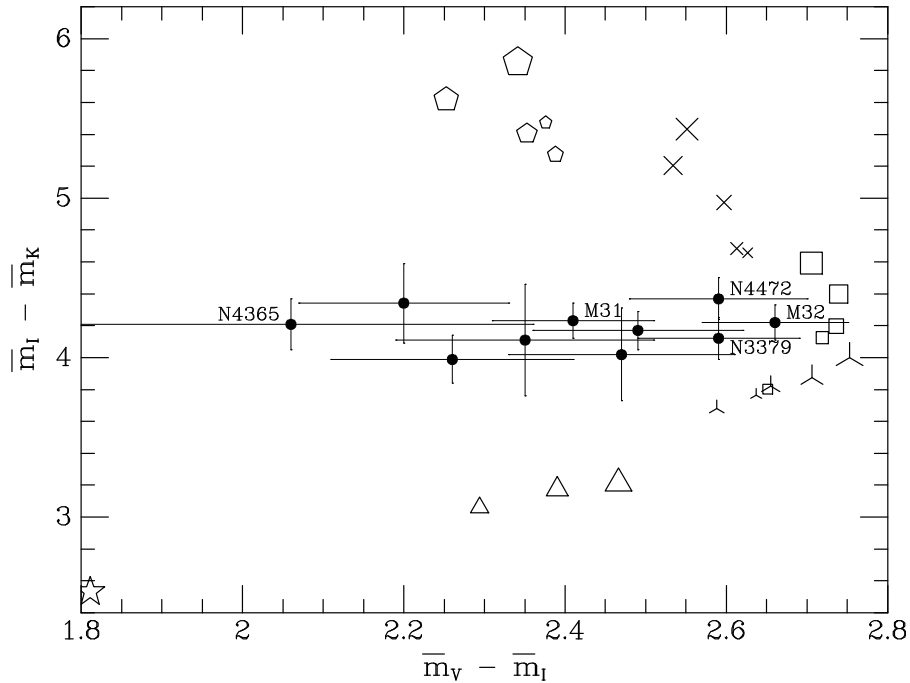}
\caption{The optical-IR SBF color $(\mibar{-}\mkbar)$ is
plotted against the purely optical $(\mvbar{-}\mibar)$.
Data for 9 ellipticals and the bulge of M31 are shown as
filled circles, with several noteworthy galaxies labeled.
The $V$ and $I$ data come from TAL90, with a few \mibar\
revisions from more recent SBF Survey observations.  
The \mkbar\ data are from Luppino \& Tonry (1993), 
Jensen \etal\ (1998a), and Pahre \& Mould (1994).
Single-burst population models from Worthey (1994) are plotted
with the symbols from that work.  The models have [Fe/H] values of
$-$1.0~(stars), $-0.5$ (open triangles), $-.25$ (skeletal triangles), 
$0.0$~(squares), +0.25 (crosses), and +0.50 (pentagons) dex.  Symbol
size is coded according to age: 3, 5, 8, 12, 17 Gyr (the more
metal poor models lack the younger ages).  Clearly, 
it is necessary to mix models of various metallicities
to reproduce the data.~~~ 
}
\label{fig:flucols}
\end{figure}

In Figure~\ref{fig:flucols}, we construct the observed
$(\mibar{-}\mkbar)$ vs.\ $(\mvbar{-}\mibar)$ SBF color-color
diagram from the available data and compare it to the
single-burst model predictions.
The comparison indicates that
ellipticals generally comprise composite stellar populations.
Only the low-luminosity, compact Local
Group elliptical M32 and,
paradoxically, NGC~4472 and NGC~3379,
the brightest ellipticals in Virgo and Leo, respectively,
approach the locus of the single-burst population models.
Further modeling would be necessary to determine the validity
of this result.

\section{The I-band SBF Distance Survey}

Along with Alan Dressler, we have completed a large survey of galaxy
distances using the $I$-band SBF method and are in the process of
finalizing the analysis.  The first ``SBF Survey'' paper \cite{sbfI}
details the data sample, observations, and calibration of the survey.
The second paper (Tonry \etal\ 1998) will present an analysis of the dynamics
of the Virgo supercluster using SBF distances.  The third paper
(Ajhar \etal\ 1998) will fully describe the theory, data reductions,
and analysis methods.  A fourth paper
(Dressler \etal\ 1998) will analyze the velocity field
around the Centaurus cluster and Great Attractor region.
Finally, we plan to make the entire imaging data set available
to the community.

We have observed
over 400 galaxies for the survey, successfully deriving distances
to about 340 within a redshift of $\sim\,$4000~km/s.  (The rest
suffered from inadequate data quality or excessive morphological
disturbance or dustiness.)
Of the early-type galaxies listed in the Third Reference Catalog of 
Bright Galaxies (de Vaucouleurs \etal\ 1991), we have observed
75\% of those within 1500~km/s and the majority out to 2800~km/s.
The sampling is fairly sparse beyond this distance.

\subsection{The Empirical Calibration}

\begin{figure}
\vspace{9.5cm}  
\caption{
\includegraphics{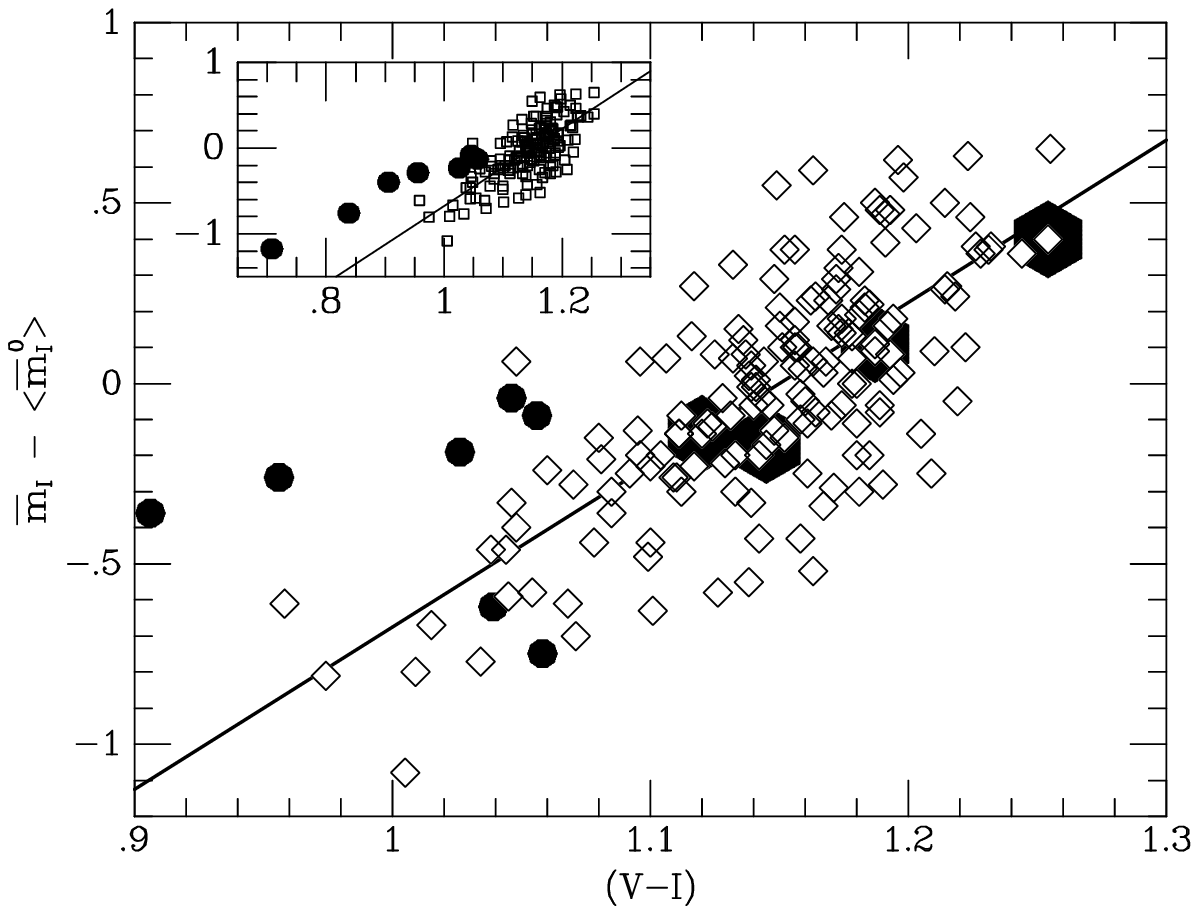} 
The empirical relation between $I$-band SBF and \viz\ color
for all galaxies belonging to the 40 SBF Survey groups.
The observables \mibar\ and \vi\ are measured for each galaxy;
\avemi\ is the value of \mibar\ in each galaxy group at the
fiducial color $\viz = 1.15$. The line represents a simultaneous 
fit to the forty values of \avemi\ along with a universal slope
for the dependence of $(\mibar{\,-\,}\avemi)$ on \viz.
Four spiral galaxies with Cepheid and SBF distances
are plotted as large, solid hexagons, demonstrating that SBF 
is the same for spiral bulges as for elliptical and S0 galaxies.
The round, solid points above the line are various locations in 
NGC~205, and those below the line are NGC~147 and NGC~185;
none of these Local Group dwarfs were used in the fit.
The inset shows the locations of NGC~5253 and IC~4182,
placed according to their Cepheid distances.
These galaxies and NGC 205 deviate from the relation
because their young blue stars change the overall galaxy color by
a large amount but are not very luminous in the $I$~band compared 
to the stars at the top of the RGB, the main contributors to \mibar.
}
\label{fig:mbar-vi}
\end{figure}

The Survey analysis uses multiple measurements of \mibar\ in 
galaxy groups and clusters to derive empirically the dependence
of \MIbar\ on \viz. SBF-I concluded that the relation is
accurately linear over the range of interest
(see Figure~\ref{fig:mbar-vi}).
Extensive comparisons in SBF-I with other distance estimators 
inspires confidence in the universality of the \MIbar--\vi\ relation.
However, an updated comparison between SBF and PNLF distances by
Mendez (1998) finds these methods no longer agree to the
extent reported by Ciardullo \etal\ (1993).

One difficulty in comparing SBF and Cepheid distances is the
fact that Cepheids are young stars residing in late-type
galaxies, while the SBF method only works well in the old stellar
populations of ellipticals and meaty spiral bulges.  We get around
this difficulty by using the group distances; the comparison 
between SBF and Cepheid group distances is especially
encouraging and is used to set the zero~point of the calibration.
The full empirical calibration derived by SBF-I is then:
 \begin{equation}
   \MIbar \;=\; (-1.74 \pm 0.07) \,+\, (4.5 \pm 0.25)\,[\viz - 1.15] \;.
	\label{eq:empcal}
 \end{equation}
This calibration is based on 10 Cepheid and 44 SBF distances 
in 7 galaxy groups.  At the fiducial color of $\viz=1.15$, it 
is 0.35 mag fainter than the Tonry (1991) calibration.  (Although
due to the steeper slope, the difference is only ${\sim\,}0.2\,$mag
at the very red colors of the Virgo giant ellipticals.)
The 1991 zero~point estimate relied solely on M31 and M32, and the 
observational and photometric errors in \mibar\ and \vi\ worked
coherently in their detrimental affect on that estimate.

Because of the much larger set of calibrating galaxies, and
the good agreement between Eq.~(\ref{eq:empcal}) and the purely
theoretical calibration of Eq.~(\ref{eq:worcal}), we have
confidence that the new empirical calibration is a good one.
It is possible that Tonry \etal\ (1998) will revise it
slightly, according to the final set of SBF~Survey measurements,
the latest Cepheid distances, and the new Galactic extinction estimates
of Schlegel \etal\ (1998), but the changes should be minor.

\subsection{The Greater Hubble Flow and $H_0$}

The SBF Survey can be tied to the Hubble flow 
via comparison with other distance estimators extending
to larger redshift.  Tying to large redshift using
the $D_n{-}\sigma$ distances of Faber \etal\ (1989) 
or the ``Mark~II'' Tully-Fisher distances (both in km/s),
SBF-I found values of the Hubble constant
near $H_0\sim86$ km/s/Mpc.  In contrast, the tie to large
redshift via Type~Ia supernova \cite{Ham95} yielded
$H_0\sim72$ km/s/Mpc.  After some discussion, SBF-I saw
no reason to exclude either large-redshift tie, and 
so offered 72--86 km/s/Mpc as the likely range of $H_0$.

It is difficult to constrain $H_0$ directly from ground-based
optical SBF measurements.  Tonry \etal\ (1998) address this
problem in the context of Virgo infall.  Section~6 reviews $H_0$
determinations from SBF measurements using \hst.  
Direct constraints on $H_0$ from the ground-based near-infrared 
SBF measurements of Jensen \etal\ (1998b) are given in Section~8.

\section{Other Ground-based Optical SBF Distances}

Several other groups have published optical SBF distances
from ground-based data.  The first of these was Lorenz \etal\ (1993)
who showed that the S0 galaxy NGC~1375 lay within
the Fornax cluster.  Tonry (1991) had derived a surprisingly bright
\mibar\ for this galaxy, placing it about~$3\;$Mpc in front of the rest
of the cluster. He suggested that the SBF measurement was corrupted
by the isophotal distortions.  Lorenz \etal\ were more meticulous
in restricting their analysis to the bulge, avoiding the
``strong disk and ring-like structure seen in absorption.''
For the neighboring elliptical NGC 1374, these authors obtained a
distance in agreement with Tonry (1991).
In addition, the \mibar\ measurements for M32 and NGC 3379 
by Sodemann \& Thomsen (1995, 1996) agree closely with those of TAL90.

A novel approach was taken by Shopbell \etal\ (1993), who measured SBF 
from digitized wide-field photographic plates of NGC 5128 (Cen$\,A$).
They obtained a distance in close agreement with the one reported by
Tonry \& Schechter (1990) using the ``traditional'' CCD approach.
Shopbell \etal\ demonstrated that, remarkably, the method could have
been applied long ago, before the development of CCDs.  However, due to
the large amount of grain noise from the photographic emulsion, it
would likely work for only the nearest ellipticals with strong SBF signals.

Simard \& Pritchet (1994) published distances to two Coma~I group
galaxies from $V$-band SBF measurements.  They found a distance
of 15~Mpc for the elliptical NGC 4494 but only 10~Mpc for the edge-on
spiral NGC 4565, 
concluding it was in the foreground.
We have $I$-band measurements from the SBF Survey that place
both galaxies at 16~Mpc.   

We are wary of $V$-band SBF because it has not been well 
characterized observationally.  Buzzoni (1993) discusses at length
the acute sensitivity of \MVbar\ to metallicity and age in his
models.  However, Worthey's (1993) models indicate that \MVbar\
is nearly as well-behaved as \MIbar.  In either case, it should
be noted that the bulge of NGC 4565 has an integrated color
similar to that of NGC 4494.
However, being so faint, the $V$-band fluctuations are difficult
to detect at $\sim\,$15~Mpc, except in giant ellipticals with ample
surface area for measuring the signal.  It would be easy to mistake
morphological distortions associated with the disk for true SBF.

Ajhar \& Tonry (1994) published SBF distances to their 19 Galactic
globular clusters, deriving the value of $\MIbar$ from RR~Lyrae
distances.
Ajhar \etal\ (1996) used that calibration along with \hst\ data
on an M31 globular to obtain 
$\mM = 24.56\pm0.12$, in agreement with Cepheids.
Finally, we note that Tiede \etal\ (1995) calculated \mvbar\ and \mibar\
for the Galactic bulge from deep star counts through Baade's window.
Their results give a Galactic center distance of $10{\,\pm\,}2\;$kpc 
with the SBF-I calibration;
most of the uncertainty is due to an uncertain \viz\ for the bulge.

\section{Ground-based SBF Measurements in the Near-Infrared}

SBF magnitudes are much brighter in the near-infrared (IR)
for early-type galaxies, with $(\overline m_I{-}\overline m_K) > 4.0$.
It therefore seems natural for the method to transit into the IR
as the detectors improve, {\it if} SBF magnitudes
behave predictably there.  Several studies of IR SBF have already
been done, and the promise held by this method has begun to
reach fruition with the thesis work of J.~Jensen.  Before surveying
the observations, we briefly examine the model predictions.

\subsection{Predicted Behavior of IR SBF Magnitudes}

Somewhere between $I$ and $J$, the sense of the \Mbar--color
relationship reverses, according to the models of Worthey (1993a).
Redder model populations therefore have brighter $JHK$ SBF magnitudes,
but the predicted trends are weak, {\it e.g.}:
$\,\overline M_K \sim -1.2\,(V{-}I)\, -\, 4.2\,$
[compare Eq.~(\ref{eq:worcal})].  On the other hand, the Buzzoni (1993) 
models still show a decline in brightness with color for \MKbar,
and are fainter by $\sim\,$0.6 mag in the mean.  Pahre \& Mould (1994)
found better agreement with the Worthey models and suggested the difference
was due to Worthey's inclusion of an empirical M-giant population.

Although the predicted changes in \MKbar\ are relatively small,
variations in metallicity and age are no longer degenerate 
in their effects, whereas they are for \MIbar.  Age differences
therefore induce scatter in the model \MKbar--metallicity relation.
For this reason, it was initially unclear whether or not \MKbar\ would
be a reliable distance indicator. 
Ironically, it has proven to be a near perfect one,
as the following section recounts.

\subsection{Observations of IR SBF}

Luppino \& Tonry (1993)
used a 256$^2$ NICMOS3 array and a $K^\prime$ filter 
($\lambda_C = 2.1\,\mu$m) to measure \mkbar\ in M31, M32, and 
Maffei~1, a heavily reddened elliptical lying close to the Galactic plane.
Adopting the Cepheid distance of 0.77~Mpc
(Freedman \& Madore 1990), they found 
$\overline M_K=-5.61\pm0.14$ for the bulge of M31 and
$\overline M_K=-5.87\pm0.14$ for M32.  Taking the M31 result
as a calibration, they derived a distance of $4.2\pm0.5\;$Mpc
to Maffei~1.  They concluded that Maffei~1 is a true giant elliptical, clearly
not a Local Group member as Spinrad \etal\ (1971) suggested, and twice
as distant as the Faber-Jackson estimate of Buta \& McCall (1983).
The 0.26~mag difference in \MKbar\ between M31 and M32 was troubling,
and they speculated that it might be due to an extended AGB in M32. 
Using M32 as the calibrator would only have increased the
Maffei~1 distance to 4.7~Mpc, however.

Pahre \& Mould (1994) also used a NICMOS3 array but a ``K-short''
filter ($\lambda_C = 2.16\,\mu$m) to measure \mkbar\ for NGC 3379 
and 8 Virgo ellipticals.  Excluding two apparent outliers, they
derived $\langle\overline M_K\rangle = -5.74$; the rms dispersion
of 0.20~mag was comparable to their typical measurement error of
0.18~mag.  They modeled the effects of a hypothetical extended
AGB on \MKbar\ and concluded that such a component must not be
common in giant ellipticals.  Their measurement for NGC 4365 indicated
that it was in the Virgo cluster proper, not in the background W~Cloud
as TAL90 had found (but see below).

Jensen \etal\ (1996) measured \mkbar\ for 7 Virgo ellipticals
and the bulge of M31 using the same instrument and filter as
Luppino \& Tonry (1993).   Combining their results with those from
Pahre \& Mould (1994), they discovered a bias affecting the low
signal-to-noise (S/N) measurements and attributed it to
errors in sky subtraction and variations in the dark current
on scales comparable to the psf.  They conservatively concluded
that $S/N \sim P_0/P_1 \gta 4$ was required for accurate SBF
measurements with present IR arrays.  Accounting for the bias,
they found $\langle\overline M_K\rangle \approx -5.62$,
with an rms dispersion of 0.29~mag, similar to the measurement errors.

More recently, Jensen \etal\ (1998a) have 
used the large format 1024$^2$ QUIRC near-IR
camera to measure \mkbar\ for 5 galaxies in the Fornax cluster, 4 in
the Eridanus group and NGC 4365.  They made several improvements to
the analysis techniques, including the use of optical images to identify
and remove globular clusters and background galaxies from the IR images. 
The improved methodology was also used to reanalyze the earlier Virgo data.
The new high-S/N observation of NGC 4365 clearly showed that this
galaxy lies about 0.65~mag behind the Virgo core towards the W~Cloud,
in precise agreement with the $I$-band distance.

Calibrating the $K$-band SBF measurements with 5 Cepheid distances,
Jensen \etal\ (1998a) derived
$\langle\overline M_K\rangle = -5.61\pm0.06$ for a sample of 11
galaxies with high-S/N data.  This is within the range of the Worthey
(1993a) models.  There is an additional systematic
uncertainty of $\sim\,$0.1 mag from the zero-point of the Cepheid scale.
No significant change in \MKbar\ was detected as a function of color 
or metallicity.  If the models are
accepted at face value, the constancy of \MKbar\ implies that bluer,
less luminous ellipticals are younger than their giant kindred.
In addition, anomalous AGB populations must be rare or absent,
though could contribute marginally to \MKbar\ in a couple 
2$\,\sigma$ outliers, including M32.  
\MKbar~is thus an excellent standard candle without need of 
color correction.

Jensen \etal\ (1998b) have pushed their ground-based 
$K$-band method further, measuring distances to NGC 4889
in the Coma cluster ($c{z} = 7186$ km/s) and NGC 3309/NGC 3311
in the Hydra cluster ($c{z} = 4054$ km/s). 
Their Coma distance translates to $H_0 = 85\pm11$ km/s/Mpc,
and their distance of 46$\,\pm\,$5 Mpc for Hydra implies
a small radial peculiar velocity for this cluster,
$v_p \lta 400$ km/s.  This is consistent with the Great Attractor 
model of Lynden-Bell \etal\ (1988), which places Hydra at a right angle
to the G.A., with the resultant motion perpendicular
to our line of sight.

The great success of the IR SBF method so far
strongly suggests that
it can be pushed even further, and attain higher accuracy,
with the larger telescopes and better IR detectors now coming on-line. 
Yet, because of the dramatically lower background,
the promise of the method is far greater with space-based
observations, as we discuss in the following section.

\section{SBF Distances from the Hubble Space Telescope}

\subsection{Mining the Archive}

A number of galaxy observation taken for other programs have been
used for SBF measurements.  Ajhar et al. (1997) have calibrated the
SBF technique with WFPC2 in the F814W filter, which approximates
the Kron-Cousins $I$~band.  They measured \mfw\ for 16 
galaxies and compared their results to \mibar\ measurements
from the ground.  The sample derives from a GTO program to study the
cores of early-type galaxies, but the integration times were sufficient
for measuring SBF amplitudes.

As discussed in Section~3.3, SBF magnitudes depend
strongly on wavelength.  For instance, an elliptical with
$(R{-}I) = 0.6$ might have $(\overline m_R{-}\overline m_I) = 1.5$.
As the total WFPC2/F814W bandpass is wider and extends a bit to the
blue of $I_{KC}$, an independent calibration was deemed necessary.
Ajhar \etal\ concluded that $\MFW \approx \MIbar$ for $\vi \approx 1.15$,
but the slope of the color dependence was steeper 
at the 2.7$\,\sigma$ level. 
The best-fit value for the slope was 6.5.  The distances obtained
with this calibration agreed well with those from the SBF Survey.

Several other authors have measured SBF distances with \hst.
Neilson \etal\ (1997) used WFPC2 observations taken in
parallel mode to measure a distance of $15.6{\,\pm\,}1.0$ Mpc
for the Virgo galaxy NGC 4478.  Thomsen \etal\ (1997) have pushed
the limits of the technique in an effort to measure the distance
to the Coma cluster elliptical NGC 4881.  The WFPC2 image of this galaxy
was taken for a study of its globular cluster system and yielded
a signal of only 0.7~$e^-/\mbar$ ({\it cf.} Section~2).
They employed a calibration based on Mg$_2$ index and arrived at
a Coma cluster distance of $102\pm14$ Mpc,
implying $H_0 = 71\pm11$ km/s/Mpc.

Morris \& Shanks (1998) measured SBF magnitudes using 3 of
the 16 WFPC2 galaxy observations in the Ajhar \etal\ (1997) sample.
Applying very different methods for galaxy subtraction, psf-fitting,
and background correction, they obtained reassuringly
similar values of \mfw\ (a median difference of 0.11 mag).
These authors interpreted their results somewhat differently, however;
we discuss their conclusions further in Section~8.1.

\subsection{SBF Programs with \hst}

Prior to refurbishment, \hst\ was not useful for SBF measurements.
Since then, however, SBF observations have been allocated time during
Cycles~5  (``The Far Field Hubble Constant'') and 6 (``The Cosmic Velocity
of the Great Attractor'') using WFPC2, and Cycle 7 using NICMOS (``The
SBF Hubble Diagram'').  We discuss each of these proposals, and a GTO
SBF study by Pahre \etal\ (1998), below.

The Cycle 5 observations were successful in providing distances to 
four Abell clusters and in calibrating the zero point of the brightest
cluster galaxy distance scale (Lauer \etal\ 1998). 
This program yielded a value of $H_0$
dependent on the Cepheid distance scale on the near end and the
linearity of the Hubble flow to 4500 km/s at the far end.  Another
goal was to test the validity of three reference frames: the CMB
frame, the Local Group frame (perhaps modified by Virgo infall), and
the ``Abell Cluster Inertial'' frame proposed by Lauer and Postman.  
The agreement between SBF distance and velocity was 
excellent in the CMB frame ($\chi^2/N = 0.3$) and poor in
the ACI frame ($\chi^2/N = 2.4$).

Pahre \etal\ (1998) have used WFPC2 IDT/GTO observations to 
measure the distance to NGC 4373, a large elliptical in its 
own group within the Hydra-Centaurus supercluster.
Adopting the calibration of Ajhar \etal\ (1997), they find
$d = 39.6\pm2.2$ Mpc for this galaxy and derive a peculiar
velocity of $415\pm300$ km/s.
This peculiar velocity is about half as large as the
838 km/s prediction from the Great Attractor model 
proposed by Lynden-Bell \etal\ (1988) based on 
$D_n{-}\sigma$ measurements.

\begin{figure}
\vspace{9.6cm}  
\caption{
\includegraphics{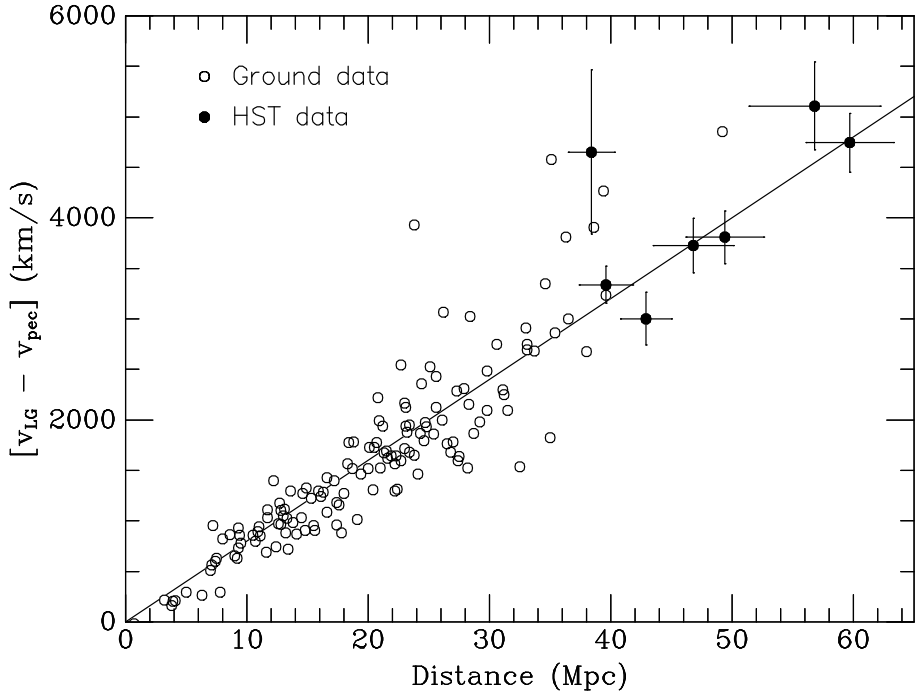} 
The $I$-band SBF Hubble diagram.  SBF distances measured
with WFPC2 on \hst\ are shown as filled symbols;
distances from the ground-based SBF Survey are shown
as open symbols, with error bars omitted for clarity.
The velocities are in the Local Group frame, corrected 
for a nonlinear Virgo infall model of amplitude 
200 km/s at the Local Group (Tonry \etal\ 1998). 
The very discrepant \hst\ point near 39~Mpc is a 
high peculiar velocity galaxy in the ``Cen-45'' cluster.
}
\label{fig:hubble}
\end{figure}

The Cycle 6 SBF program explores this issue further,
probing the size of the Great Attractor by
measuring SBF distances (and peculiar velocities)
to two galaxies in the Centaurus cluster, two galaxies at 5000 km/s 
judged to be on the far side of the G.A.\ (and hence should
reveal backside infall), and one galaxy of comparable distance at an
angle where the peculiar velocity should be negligible.  Because of
the NICMOS cryogen problem, the observations for this program are 
not yet complete, but the data taken so far appear excellent.
We expect them to provide
definitive answers to the questions of where the 
Centaurus/G.A.\ flow comes to rest with respect to the CMB,
and what the precise amplitude of the flow is.
Figure~\ref{fig:hubble} displays the SBF Hubble diagram 
for the Cycles~5 and~6 data, combined with the
ground-based $I$-band Survey data.

SBF is about 30 times brighter in the $H$-band than in the $I$-band,
and NICMOS affords an enormous advantage over ground-based IR observations
because its sky background is at least 100 times fainter.  The
Cycle 7 program seeks to calibrate SBF in the F160W filter (similar to $H$)
and then measure $H_0$ at three different distances: 4500 km/s,
7000 km/s, and 10,000 km/s.  The project has collected data on a
``calibrator'' sample consisting of 10 nearby galaxies in clusters where
good Cepheid distances exist (Leo~I, Virgo, and Fornax).  These
galaxies were chosen to span a range of color and luminosity that 
will allow the dependence of F160W SBF on metallicity and age
to be characterized.

The ``nearby'' sample at 4500 km/s consists of 5 galaxies for which 
there are WFPC2 observations, making it possible to directly tie together
the SBF distances measured by the two instruments.
Along with the calibrators,
these should yield a Hubble constant valid within about 4500 km/s.
The ``intermediate'' sample comprises 6 central galaxies in Abell clusters
in the redshift range 5500--9500 km/s.  
These clusters were selected to be at the vertices of an octahedron
which symmetrically straddles the Galactic plane.  We thus expect that
the mean $H_0$ derived from this data set will
be extremely insensitive to the velocity reference frame adopted.
Moreover, we will obtain a good estimate of the reference frame 
in which these galaxies are at rest. 
Finally, the ``distant'' sample was trimmed to a single cluster 
at 10,000 km/s.  Most of the data for this project have been taken, 
and the first pass reductions are extremely promising.

\section{Evaluating SBF as a Distance Indicator}

\subsection{Addressing some criticisms\label{subsec:crits}}

One of the primary points of contention surrounding the SBF method is 
its calibration.  
The present calibration from SBF-I still may not be perfect. 
Systematic errors of ${\sim\,}0.1\,$mag could remain, but we have noted
reasons for believing that the zero-point error is not much larger than this.
Moreover, earlier indications of residual correlations between inferred
SBF distance and galaxy luminosity (Tonry \etal\ 1989), integrated
color \cite{TAL90}, and Mg$_2$ index (Lorenz \etal\ 1993; Tammann 1992)
vanish in the light of the new \MIbar--\vi\ calibration \cite{sbfI}.

Recently, 
Morris \& Shanks (1998, hereafter MS98) have suggested that the
lower limit for errors in $I$-band SBF distances is actually 0.17~mag,
based on their reductions of three galaxy observations in
the \hst\ archive.
Much of this conclusion appears to stem from the 0.05 mag
uncertainties they derived for their \vi\ values.
If \viz\ is known to only 0.05 mag, then
the minimum SBF distance error is actually 0.23~mag, due to
the $4.5\,(V{-}I)_0$ term in the \MIbar\ calibration.
For this reason, SBF-I expended much effort to ensure accurate
and uniform photometry to better than 0.02 mag in \vi\
for the SBF Survey (see 8.3.4 below).

All three of the \hst\ observations analyzed by MS98 were included
in the sample of 16 galaxies from Ajhar \etal\ (1997).
Unlike MS98, Ajhar \etal\ did not attempt to analyze
data from the WF~cameras, which badly under\-sample the psf.
Comparing just the MS98 PC chip measurements from their Table~8
to Ajhar \etal\ gives differences of
$+0.21\pm0.11\;$mag for NGC~3379,
$-0.11\pm0.08$ mag for NGC~4406, and
$-0.07\pm0.11$ mag for NGC~4472.
The agreement in \mbar\ itself is reasonable;
only the NGC 3379 results differ by nearly $2\,\sigma$.
Pahre \etal\ (1998) also measured \mbar\ for the same
NGC 3379 and NGC 4406 \hst\ data in order to test 
their SBF analysis method on the marginally sampled PC images.
Compared to the \mbar\ values of Pahre \etal, those of MS98 differ
by $+$0.36 mag for NGC 3379 and $+$0.04 mag for NGC 4406.
Both Ajhar \etal\ and Pahre \etal\ found results for NGC~3379 
consistent with the ground-based numbers from the SBF survey.~~  

The MS98 measurement for NGC 3379 is thus inconsistent with
both the Ajhar \etal\ and Pahre \etal\ \hst\ results, and with the
ground-based results from the SBF Survey and Sodemann \& Thomsen (1996).
One cannot conclude from this single discrepancy,
or from \vi\ measurements with 0.05 mag uncertainties,
that the average distance error in the hundreds of ground-based
SBF distances is 0.25~mag and that the minimum error is 0.17~mag. 
We refer the reader to SBF-I for an extensive statistical analysis
of the ground-based SBF distance errors.~ 

\subsection{``Cosmic'' Scatter}

We rehash here the evidence for universality in the behavior of the
two main SBF magnitudes utilized for estimating galaxy distances.

Based on SBF and color measurements for $\sim\,$150 galaxies in
$\sim\,$40 nearby galaxy groups, SBF-I concluded that the 
quantity $\MIbar - 4.5\,[\viz-1.15]$ is a standard candle
among early-type galaxies in the color range $1.0<\viz<1.3$.  
Calibration via Cepheids yielded an absolute magnitude
of $-1.74$ mag for this standard candle.  From 
an analysis of $\chi^2$, they concluded that the intrinsic,
or ``cosmic,'' scatter was less than 0.1$\,$mag; most likely
it was $\sim\,$0.05 mag.
The stellar population models of Worthey (1993) indicate
that the above \MIbar--\viz\ relation is a standard candle
with intrinsic scatter $<\,$0.11 mag, depending upon
the amount of variation present among the stellar populations
of elliptical galaxies.  These models give a calibration
brighter by $\sim\,$0.08 mag than the empirical one.

The second most commonly utilized SBF magnitude for distance
estimation is \mkbar. The observations by Jensen \etal\ (1998a) 
indicate that \MKbar\ is by itself a standard candle with a
cosmic scatter of only 0.06 mag for early-type galaxies
in the limited color range $1.15<\viz<1.27$.  The models predict
that \MKbar\ should systematically brighten by $\sim\,$0.15 mag 
even over this color range, if all ellipticals are coeval.
Thus, if both the models and the observations are correct, there
must be an age-metallicity conspiracy among early-type galaxies
to keep \MKbar\ constant.  Further investigation along both lines
is needed to test the significance of this result.

\subsection{What can go wrong} 

If SBF is such a great standard candle, why might some
distances be wrong?  Below, we list possible problems
that can affect distance estimates.~~

\subsubsection{Difficult Galaxies}

There have been some clear discrepancies with SBF distances
involving measurements on non-ideal galaxies, such as the
edge-on spiral NGC 4565 (Simard \& Pritchet 1994, as 
compared to the SBF Survey) and the disky~S0 NGC 1375 
(Tonry 1991, compared to Lorenz \etal\ 1993).  For these two
cases it appears that the better measurement gives the correct
distance.  Thus, the problems were with the reductions, not due to
intrinsic difference in SBF magnitude between spiral bulges and
ellipticals. It pays to take pains.
With well over 300 SBF distances in the Survey, it is difficult
to ensure a uniform pain threshold for all the reductions; there
may be a few bad distances in the complete data set for this reason.

Unfortunately, Cepheids only dwell in the most difficult galaxies for SBF.
A case in point is the flocculent spiral NGC 7331, for which we
attempted to measure an SBF distance from its smooth outer disk.  
SBF-I reported 12~Mpc for this galaxy
in anticipation of the Cepheid distance, which comes in
at 15$\,\pm\,$1 Mpc (S.\ Hughes 1998, priv$.\,$comm.).
We reanalyzed the data, but they gave the same \mibar.  
Although the outer disk appears smooth, 
the stars within it must be correlated on the
scale at which the SBF was measured,
$\sim\,$22~pc at this distance;
if this is the case, the SBF method will not work.
Fortunately, this is not a problem for ellipticals 
and other ``hot'' stellar systems.
It thus appears we were over-zealous in our attempt to measure
SBF distances for every possible Cepheid-bearing galaxy.~~

A similar problem may affect the SBF distance for the edge-on
disk galaxy NGC~3115.  The PNLF and RGB tip methods give
$10.9\pm0.7$ Mpc (Elson 1997), while the SBF distance is
$9.2\pm0.5$ Mpc. Remeasurement of \mbar\ from Elson's \hst\ data
in a clean region of the bulge well away from the disk would
help in understanding the discrepancy for this galaxy.

\subsubsection{PSF Mismatch}

An accurate SBF amplitude depends on having a good star to serve
as a psf template.
This is usually not a problem, but occasionally for a galaxy at high
galactic latitude, the pickings get rather slim.  As all power
spectrum measurements are referenced to the psf template
power spectrum, an error
of 5\% in its normalization translates into a 0.05 mag error 
in \mbar.  This is a bigger problem in the IR, where psf stars must
contend with an extremely bright sky (see Jensen \etal\ 1996).
For \hst\ images, one has recourse to a synthetic psf,
although most groups opt for
empirical ones if at all possible (e.g., Ajhar \etal\ 1997;
Pahre \etal\ 1998; Lauer \etal\ 1998).

\subsubsection{Bad Backgound Luminosity Function Model}

The ability to detect, remove, and model the faint globular
clusters and background sources so that $P_0$ is dominated by
the SBF and not by the background variance $P_r$,
and so that $P_r$ can be accurately estimated,
is the limiting factor in the ground-based $I$-band SBF method.
Greatly improved background source removal due to superior resolution
is the big advantage \hst\ holds for the optical SBF method.
(The major advantage of \hst\ in the IR is the much lower background.)

Faint point source removal and luminosity function modeling
for estimating $P_r$ are discussed in detail by Tonry \& Schneider (1988),
TAL90, and Ajhar \etal\ (1998), to which we refer the reader.
One thing worth noting here is that the uncertainty estimate
for $P_r$ should not be made directly proportional to the $P_r$ estimate.
Otherwise, if the estimated $P_r$ is less than the true residual
variance, the $P_r$ uncertainty will be underestimated
by the same factor, making the derived \mbar\ simultaneously
too bright and overly significant.  This is avoided by estimating
the uncertainty in $P_r$ from the depth of the point source 
removal (see Ajhar \etal\ 1998).

\subsubsection{Good \mibar, Bad \vi}

With all the trouble involved in measuring \mibar, one might think
something as simple as the galaxy color would be easy.
This is a dangerous trap.  Since 
$\MIbar \sim {4.5\,}(V{-}I)$, the \viz\ color must be known to
0.024~mag for an accuracy of 5\% in distance.
SBF-I described an entire secondary survey undertaken 
on the McGraw-Hill 1.3~m telescope to help ensure adequate
\vi\ surface photometry for the primary SBF Survey.

If a color gradient is present in a galaxy, care must be taken to 
calculate \MIbar\ using the \viz\ color determined from the region
over which \mibar\ was measured.  There are bulge/disk concerns
here, too.  Galaxy disks are usually bluer than their bulges;
if \mibar\ is measured in the bulge, then the disk must be entirely
removed before measuring \vi, or the \MIbar\ estimate will be too
bright.  Of course, if dust contaminates the color measurement, 
the \MIbar\ estimate will be too faint.  (SBF-I discuss these
issues further.)

\subsubsection{Bad Extinction Estimate}

Finally, even if one is careful about photometry and color gradients,
an inaccurate extinction estimate will produce a bad distance.
Because of the way \viz\ comes into the calculation of \MIbar, 
underestimating the Galactic extinction actually yields a 
{\it smaller} distance, as the effect of making \mibar\ artificially
dim is overwhelmed by the effect of making \viz\ too red.
The error introduced into the distance modulus from an error 
$\delta A_B$ in the $B$-band extinction is:
 $  \,\delta(\MIbar{-}\mibar) \,\approx +0.83\,{{\delta}A_B}\,.$
SBF-I used Burstein \& Heiles (1984) extinctions; in the 
next paper we will convert to the Schlegel \etal\ (1998) extinctions
determined from 100$\,\mu$m dust emission.

\section{The Future of SBF}

To push much further with optical SBF distances from
the ground will require adaptive optics techniques to
deliver images with seeing consistently $\lta\,$0\farcs45 FWHM.
This may soon be a reality, but the combination of
the 0\farcs1 psf and a sky darker by orders of magnitude
places the real future of SBF observations in space.

It seems likely to us that further application of the 
optical SBF method from space
will eventually map out the pattern of
galaxy and cluster velocity flows to distances at which 
the motions become just a few percent of the Hubble velocity.  
There is no other method for measuring elliptical galaxy distances
that can compare for completeness, depth, and accuracy.~~~ 

Before this can be accomplished, the slope of the SBF calibration
in the F814W filter must be verified directly from multiple
measurements of galaxies in tight groups and clusters.  This will
avoid compounding errors from individual ground-based distances.
With recent F814W observations of several ellipticals in Fornax,
there is now enough data in the \hst\ archive to do this,
and so we expect to have a final calibration soon.

With SBF being so much brighter and so well-behaved
in the near-IR, and the gain in
terms of a darker background being so immense, IR SBF from 
space may be the most promising method of all.  Although most of
this review has been concerned with optical measurements, it is
possible that a similar review in another decade will deal 
almost exclusively with IR SBF measurements.  Again, 
there needs to be more done to ensure accurate calibrations
of \MKbar\ and \MHbar.

Finally, there is need for improved models, and will
remain such need until we fully understand the behavior
of \Mbar\ across the spectrum.  
The SBF method will then be independent of the Cepheid calibration.
Moreover, we will have an excellent handle on the age and
metallicity mixtures in elliptical galaxies, and how these
mixtures change with luminosity and other 
\hbox{galaxy properties.}~

{}

\end{document}